\documentclass[aps,prb,twocolumn,groupedaddress,showpacs]{revtex4}
\usepackage{epsfig}
\usepackage[dvipsnames,usenames]{color}
\emergencystretch=\maxdimen
\begin{document}

\title{Disorder Line and Incommensurate Floating Phases in the Quantum Ising 
Model on an Anisotropic Triangular Lattice}

\author{V.I. Iglovikov}
\affiliation{Physics Department, University of California, Davis,
California 95616, USA}

\author{R.T. Scalettar}
\affiliation{Physics Department, University of California, Davis,
California 95616, USA}

\author{J. Oitmaa}
\affiliation{School of Physics, The University of New South Wales,
Sydney 2052, Australia}

\author{R.R.P.~Singh}
\affiliation{Physics Department, University of California, Davis,
California 95616, USA}

\begin{abstract}
We present a Quantum Monte Carlo study of the Ising model in a
transverse field on a square lattice with nearest-neighbor antiferromagnetic exchange interaction J
and one diagonal second-neighbor interaction $J'$, interpolating between
square-lattice ($J'=0$) and triangular-lattice ($J'=J$) limits. At a
transverse-field of $B_x=J$, the disorder-line first introduced by
Stephenson, where the correlations go from Neel to incommensurate, meets
the zero temperature axis at $J'\approx 0.7 J$.   Strong evidence is
provided that the incommensurate phase at larger $J'$, at finite
temperatures, is a floating phase with power-law decaying correlations.
We sketch a general phase-diagram for such a system and discuss how our
work connects with the previous Quantum Monte Carlo work by Isakov and
Moessner for the isotropic triangular lattice ($J'=J$).  For the  isotropic triangular-lattice,
we also obtain the entropy function and constant entropy contours
using a mix of Quantum Monte Carlo, high-temperature series expansions
and high-field expansion methods and show that phase transitions in the
model in presence of a transverse field occur at very low entropy.
\end{abstract}

\pacs{
05.10.-a, %% Computational methods in statistical physics and nonlinear dyn
05.30.Rt, %% Quantum phase transitions 
75.10.Jm, %% Quantized spin models, including quantum spin frustration
75.40.Mg, %% Numerical simulation studies
}
\maketitle

%%%%%%%%%%%%%%%%%%%%%%%%%%%%%%%%%%%%%%%%%%%%%%%%%%%%%%%%%%%%%%%%%%
\section{Introduction}
%%%%%%%%%%%%%%%%%%%%%%%%%%%%%%%%%%%%%%%%%%%%%%%%%%%%%%%%%%%%%%%%%%

The Ising model in a transverse magnetic field, 
\begin{eqnarray}
\hat H = + \sum_{i,j} J_{ij} \hat S_i^z \hat S_j^z 
- B_x \sum_i \hat S_i^x,
\end{eqnarray}
illustrates a variety of interesting statistical mechanics behaviors in
part because of the simplicity of its mapping to an equivalent classical
problem in one higher
dimension\cite{schultz64,pfeuty70,suzuki71,stinchcombe73,blote02,sachdev-book}.
In the case when the exchange coupling $J_{ij}<0$ is ferromagnetic, the
model exhibits a quantum phase transition: increasing $B_x$ causes the
emergence of a paramagnetic phase at $T=0$.  On the other hand, on a
triangular lattice when $J_{ij}>0$ is antiferromagnetic, $B_x$ can have
the opposite effect and cause order to occur by
removing\cite{chandra-sondhi-moessner} the large ground state degeneracy
($s(T=0) \approx 0.32$) present in the zero field case\cite{wannier50}.
The antiferromagnetic transverse field model on the isotropic triangular lattice was
studied by Isakov and Moessner \cite{isakov03} using Quantum Monte Carlo
simulations, who concluded that there are three different phases: a
paramagnet and two distinct ordered phases distinguished by the relative
dominance of quantum or thermal fluctuations.

Experimental motivation for the study of the transverse field Ising
model dates back to deGennes\cite{degennes63}, who considered the
ferroelectric KH$_2$PO$_4$ in which a double well structure of the
proton position corresponds to the Ising variable, and the transverse
field represents inter-well tunneling\cite{stinchcombenote}.  Much more
recently, transverse field Ising models have also begun to be realized
experimentally in cold atom systems and this provides the immediate
motivation for our work.  The Maryland group\cite{monroe10} has
assembled small, highly-connected clusters of trapped $^{171}$Yb$^{+}$
ions, and demonstrated a sharp crossover from paramagnetic to
ferromagnetic behavior as the Ising coupling is scaled up relative to
the transverse field.  Similarly, the Bollinger group at
NIST\cite{britton12} is exploring larger collections of up to hundreds
of Be ions in a Penning trap in triangular geometries with a
spin-dependent optical dipole force with an adjustable power law decay.
Transverse field Ising models with dipolar interactions have also been
considered\cite{tabei08} in the context of solid state systems such as
the glassy low temperature properties of
LiHo$_x$Y$_{1-x}$F$_4$.\cite{LiHoYF}	

In this paper we calculate the thermodynamics and phase transitions in
several specific instances of the transverse field Ising model in two
spatial dimensions, using a combination of high temperature and high
field series expansions\cite{book,series-review} and continuous time
Quantum Monte Carlo (CTQMC) approaches\cite{blote02,isakov03,sandvik}. Our
motivations are twofold.  First, in light of the developments in
cold-atom systems, we study the entropy function of the triangular
lattice model with ferromagnetic and then antiferromagnetic couplings.
In both cases CTQMC results are compared with high temperature and high
field expansions.  When $J>0$, a key conclusion is that the ordered
phases which are induced by $B_x \neq 0$ occur at very low entropy per
particle $s$.
For these transitions to be accessible experimentally, $s \lesssim 0.1
\, k_{\rm B}$ will typically be required.  Longer range couplings,
present in the experimental systems described above, are shown not to
dramatically alter the isentropes.  In particular, ordered phases, if
present, will still occur only at rather low entropy.

Our second motivation is to understand how the Ising-like Neel
transition in an unfrustrated system gives way to incommensurate order
and Kosterlitz-Thouless type behavior as frustrating further neighbor
interactions are added.  For this, we consider a square lattice model with
nearest neighbor interaction $J$ and an antiferromagnetic next-nearest-neighbor interaction $J'$ along one
diagonal, which then interpolates between the square and triangular
lattice limits.  There have been many previous studies of
Heisenberg and Hubbard models in this geometry \cite{zheng99}, but we
are not aware of any previous work on the quantum Ising model except in
the weakly frustrated case \cite{selke}.  Indeed, this model raises
several challenging questions for any numerical study, but also,
as we shall demonstrate, some interesting and novel physics.

For the classical Ising model ($B_x=0$) it is
known\cite{stephenson70,eggarter75} that there is a conventional second
order phase transition for $J' \textless J$, but that $T_c$ vanishes for
$J' \geq J$.  As $J' \rightarrow J$, the critical temperature
vanishes as $T_c \sim 2 (J-J')/{\rm ln}2$. The order in the low
temperature phase remains commensurate in the two-sublattice pattern for
all $J'/J<1$. Stephenson introduced a disorder line in the $T/J$, $J'/J$
parameter space, where the short-range order in the system moves away
from ($\pi,\pi$). Above $J'/J=0.6$, the disorder line comes very close
to the phase transition line, but the phase transition remains pinned to
the ($\pi,\pi$) state.  We would like to understand the fate of the
disorder line in presence of the transverse-field $B_x$ and see if there
is a resulting incommensurate order of the Kosterlitz-Thouless (KT) type
as would be expected for a system with an emergent phase variable with XY symmetry as
discussed by Bak and Villain \cite{bak}.

In the triangular-lattice limit studied by Isakov and Moessner, the
order is locked to a commensurate 3-sublattice pattern. This gives rise
to $6$th order anisotropies for the emergent phase variable.  Although
such an anisotropy is irrelevant at the finite temperature phase
transition from the paramagnetic phase, thus giving rise to a KT
transition with a power-law phase, it ultimately succeeds in driving the
system to long-range order, hence their conclusion of $3$ different
phases in the model. In the $T=0$ limit, our model is equivalent to a
$3$-dimensional system and therefore can have a Lifshitz point and a true
incommensurate long-range ordered phase.\cite{bak}  However, there can not be such
a long-range ordered phase at finite temperatures.  We have not studied
the very low temperature limit of our model, where the system may, for
some ranges of $J'/J$, lock in to different commensurate phases, thus
giving rise to long-range order which could extend to some finite but
very low temperatures. Our system
has some similarities with the well studied ANNNI
models \cite{selke88} as far as the onset of incommensurate order is
concerned. However, one key difference from the ANNNI models is that
in the latter one particular direction is singled out by the interactions and becomes the direction
in which incommensurate order and various commensurate lock-ins occur. In
contrast, in our model incommensurate wave vectors and commensurate phases
can have a much more two-dimensional character.

The remainder of this paper is organized as follows.  In Sec.~II we
summarize our three calculational approaches, Quantum Monte Carlo and
high temperature and high field expansions.  In Sec.~III we discuss the
square to triangular interpolation and the nature of the phases and
phase transitions.  Although this is the second of the motivations
presented above, we discuss it first, since it illustrates some of the
rich physics introduced by frustrating interactions.  We defer to Sec.~IV
the thermodynamics of the triangular lattice antiferromagnetic
Ising model of relevance to optical lattice experiments.
Sec.~V summarizes our results.

%%%%%%%%%%%%%%%%%%%%%%%%%%%%%%%%%%%%%%%%%%%%%%%%%%%%%%%%%%%%%%%%%%
\section{Calculational Approaches}
%%%%%%%%%%%%%%%%%%%%%%%%%%%%%%%%%%%%%%%%%%%%%%%%%%%%%%%%%%%%%%%%%%

\subsection{Quantum Monte Carlo Method}

We employ the CTQMC algorithm described in [\onlinecite{isakov03}].  In
this method, we use the Suzuki-Trotter formalism to map a $2D$ quantum
Hamiltonian, which is on a triangular lattice, onto the $3D$ classical
Ising model on a stacked triangular lattice. The number of
layers $L_\tau$ in the extra dimension is $L_\tau = \beta / \epsilon$, where
$\beta$ represents the inverse temperature and $\epsilon$ the
discretization step. The mapping becomes exact as $\epsilon \to 0$.
In this limit, the number of layers becomes infinite and computationally
much less intractable. However, an alternate view as $\epsilon \to 0$
is to think of consecutive spins, which have the same value
along the extra dimension, as parts of continuous segments rather than
living on 
individual, discrete lattice points. We treat these segments as our dynamical
objects in the simulation. This approach makes the simulation
algorithm non-local, memory efficient and allows one to work explicitly
in the $\epsilon \to 0$ limit.\cite{isakov03,reiger}

Measured quantities include the real space spin correlations 
$C({\bf r})$,
\begin{eqnarray}
C({\bf r}) = \langle \hat S_i^z \hat S_{i+{\bf r}}^z \rangle \,\,\,,
\label{Eq:Cr}
\end{eqnarray}
and their momentum space counterparts, the magnetic structure factor
$S({\bf q})$,
\begin{eqnarray}
S({\bf q}) = \frac 1 N \sum_{\bf r} 
e^{i {\bf q} \cdot {\bf r}} C({\bf r}) \,\,\,.
\label{Eq:Sq}
\end{eqnarray}
We also extract the Binder ratio,
\begin{eqnarray}
U_L =  1 - \frac{\langle M^4 \rangle}{3 \langle M^2 \rangle^2 }
\label{Eq:UL}
\end{eqnarray}
where $M^2$ is the value of the structure factor at the ordering
wave vector.

Entropy $s$ is measured by numerical integration of the internal
energy $\langle E \rangle$ 
\begin{eqnarray}
s(T) = {\rm ln} \, 2  + \frac{\langle \, E(T) \, \rangle}{T}
- \int_T^{\infty} \, dT' \, \frac{\langle E(T') \rangle}{T'^2}
\label{Eq:ST}
\end{eqnarray}

We have checked that our algorithm leads to data in complete agreement
with exact diagonalisation results on the $3 \times 3$ lattice, and
other limiting cases, for which analytic solutions are possible.

We imposed periodic boundary conditions and did simulations for
different lattice sizes $L$ and different sets of the magnetic field
$B_x$ and the coupling constant $J_{ij}$. For each simulation, we
estimated the autocorrelation time $\tau$ and performed $5000 \tau$
Monte Carlo sweeps for thermalisation and $10^5 \tau$ to carry out the
measurements.  Results for the ferromagnetic case were on 30x30
lattices, and for the antiferromagnetic
case on 27x27 lattices, unless otherwise stated.

\begin{figure}[t] 
\epsfig{figure=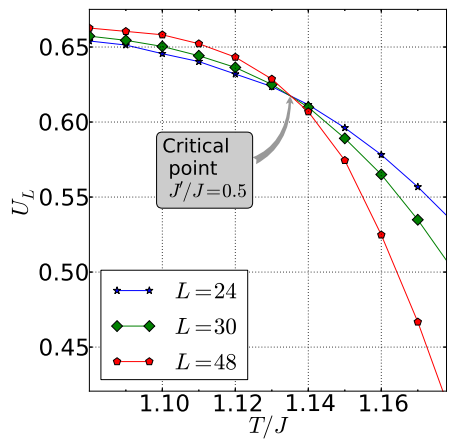,height=6.0cm,width=6.0cm,angle=0,clip}
\caption{(Color online)  
Binder crossing for $J'/J=0.5$ at $B_x/J=1$.
For $J'/J$=0 we have the square lattice $T_c/J=2.269$.  
Here the critical temperature has been suppressed by a factor of
two by the frustration introduced by the next-near neighbor
interaction $J'$ and by the quantum fluctuations due to $B_x$.
\label{Fig:bindercrossBx1plane}
}
\end{figure}

\subsection{High Temperature Series Expansions}

High temperature series expansions are based on a Taylor series
expansion for the Boltzmann factor
\begin{equation}
\exp{(-\beta \hat H)}= \sum_n {(-\beta)^n\over n\\!} \hat H^n
\end{equation}
The trace of $\hat H^n \hat O$, where $\hat O$ is some local operator
involves only a finite number of sites and hence can be evaluated, up to
some order, by a straightforward though cumbersome method.  A
particularly efficient way of calculating high order series expansions
for various extensive and intensive properties of the model in powers of
$\beta$ is the Linked Cluster Method. The details of the technique can be
found in the literature.\cite{book,series-review} We have used this
approach to obtain thermodynamic properties of the model for the
triangular-lattice transverse-field Ising model with nearest neighbor
and second neighbor interactions.

\subsection{High Field Expansions}

Since the field term of the Hamiltonian is exactly soluble, one can also
develop a series expansion in powers of $J/B_x$ for the ground state
properties of the model. We have not done such a high order expansion at
$T=0$. However, we have used finite temperature, high field expansions
to order $(J/B_x)^2$, to calculate entropy and other thermodynamic
properties.  These are useful in determining the isentropic contours at
low temperatures and high fields.

%%%%%%%%%%%%%%%%%%%%%%%%%%%%%%%%%%%%%%%%%%%%%%%%%%%%%%%%%%%%%%%%%%
\section{Square to Triangular Lattice Interpolation}
%%%%%%%%%%%%%%%%%%%%%%%%%%%%%%%%%%%%%%%%%%%%%%%%%%%%%%%%%%%%%%%%%%

\begin{figure}[t] 
\epsfig{figure=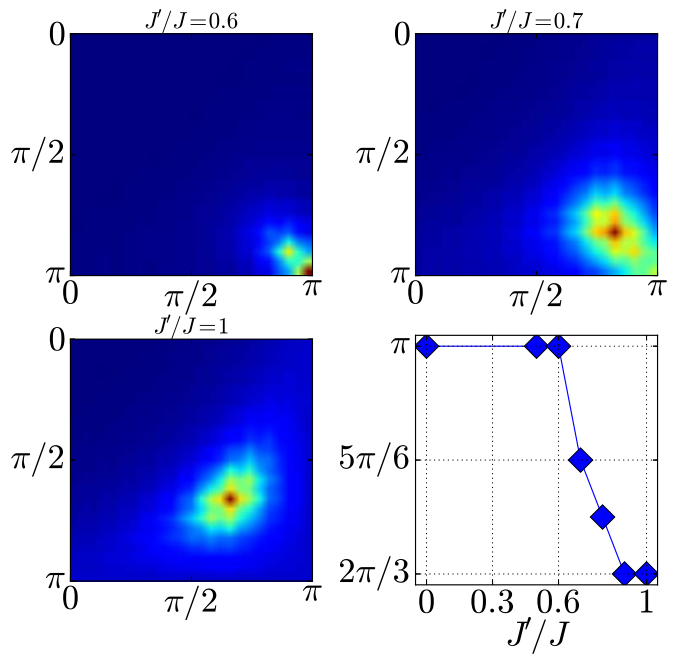,height=9.0cm,width=9.0cm,angle=0,clip}
\caption{(Color online)  
Evolution of the static structure factor with increasing $J'/ J$ (ie
as the triangular lattice is approached) for $B_x=1, T = 1$.
For $J'/J < 0.7$ the structure factor is peaked at
the AF wavevector ${\bf q}=(\pi,\pi)$.  The beginning of a shift away
from this value is evident at $J'/J=0.7$ and completely
unambiguous at $J'/J=1.0$, where the peak is at the expected triangular
lattice value $(2\pi/3,2\pi/3)$.
\label{Fig:Sqshift}
}
\end{figure}

\begin{figure}[h] 
\epsfig{figure=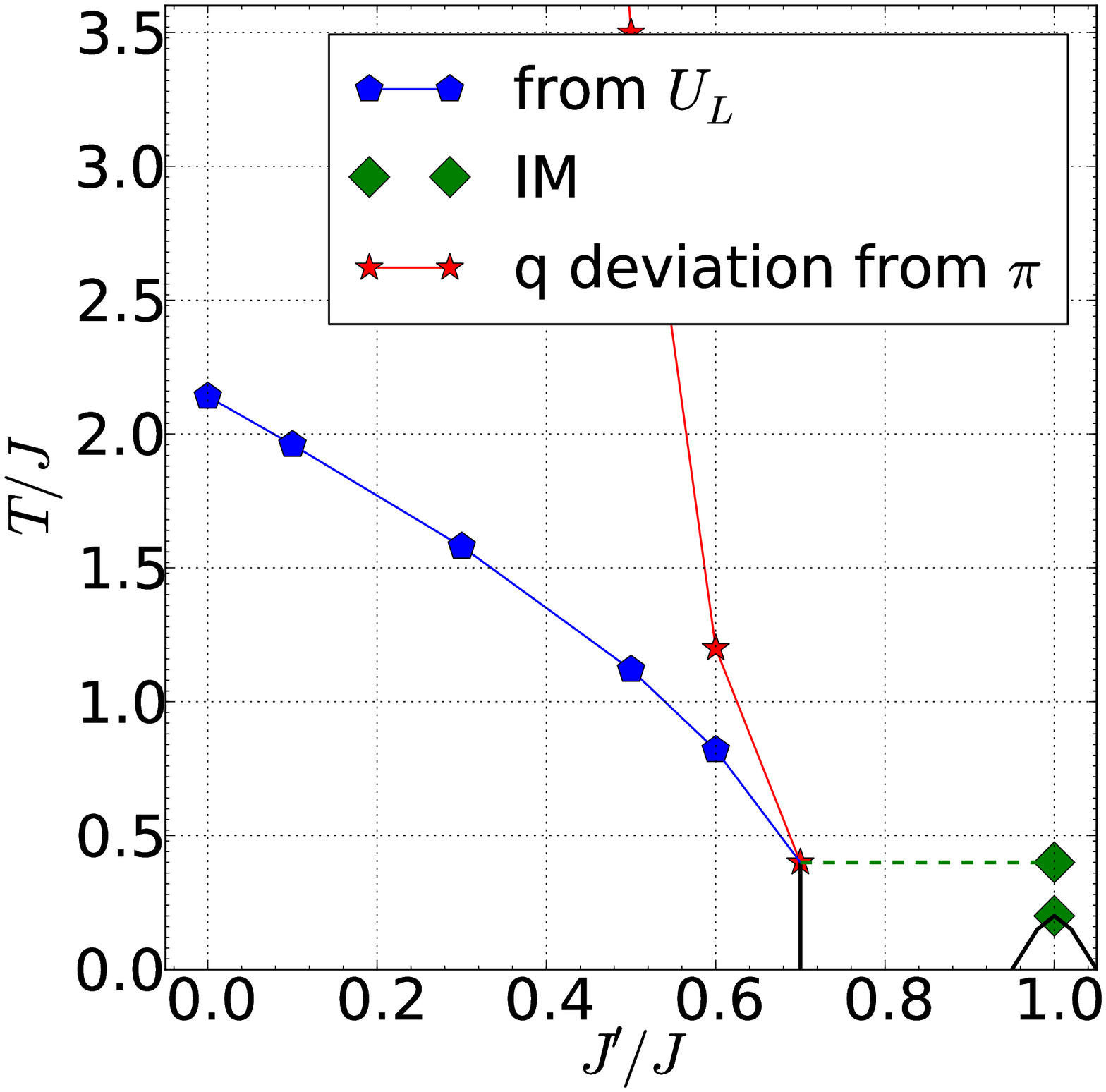,height=6.0cm,width=6.0cm,angle=0,clip}
\caption{(Color online) Phase diagram in the $T/ J$ vs $J'/J$ plane.
Pentagons are critical points representing a second order phase
transition. Stars denote the disorder line at which
the ordering wavevector shifts from $(\pi,\pi)$. 
Diamonds represent critical
temperatures obtained by Isakov and Moessner in the
triangular lattice limit.  The dashed line separates a KT
phase from a paramagnetic phase.
\label{Fig:conjecturedphasediag}
}
\end{figure}

In the introduction we briefly reviewed the physics of the square to
triangular lattice interpolation of the classical antiferromagnetic
Ising model.  In this section, we generalize this problem to the case
when a transverse field $B_x$ is included, and connect to what is known
about the problem with $B_x \neq 0$ in the triangular and square lattice
limits.  We consider a square lattice geometry with near-neighbor
coupling $J$ and with an additional next-near-neighbor interaction  $J'$
across one of the diagonals of each plaquette.  In the limit $J'=J$ the
triangular lattice is realized.  As mentioned in the introduction, the
nature of the phase diagram in the three dimensional space of $T/J,
B_x/J, J'/J$ is already understood in certain limiting cases.

In the classical limit $B_x=0$, Stephenson \cite{stephenson70} 
solved the model analytically and calculated various correlation functions.  Because
the square lattic is bipartite, without loss of generality one may
restrict consideration to a ferromagnetic choice $J<0$. The critical
temperature as a function of $J'/J$ is obtained from the transcendental
equation\cite{eggarter75}
\begin{eqnarray} 
t' = \frac{t^2 + 2 t - 1}{t^2 - 2 t -1}
\end{eqnarray} 
where $t'={\rm tanh}\,{(J'/T)}$ and where
$t={\rm tanh}\,{(J/T)}$.  This equation has solutions for which $T_c$
increases from the square lattice value $T_c/J=2.269$ for additional
ferromagnetic coupling $J' < 0$, and for which $T_c$ decreases for
antiferromagnetic $J'>0$.  The critical temperature vanishes in the
triangular limit $J'=-J$ and remains zero thereafter.

Similarly, in the $J'/J=0$ plane, the square lattice has a second order
phase transition with $T_c=2.269$ at $B_x=0$, and a $T_c$ which
decreases as $B_x$ grows, ending in a Quantum Critical Point at $T=0,
B_x/J=3.05$\cite{book}.  Finally, in the $J'/J=1$ plane
(antiferromagnetic Ising model in a triangular lattice in a transverse
field) Isakov and Moessner argued\cite{isakov03} that turning on $B_x$
induced an order-from-disorder transition in which there are in fact two
distinct (Kosterlitz-Thouless and clock) ordered phases.  The maximal
critical temperature is $T_c/J \approx 0.4$ at transverse field $B_x/J
\approx 0.8$.  Their study motivates us to pick a fixed value of
$B_x/J=1$ near this maximum to provide a detailed description of the
evolution from square to triangular geometry.

Figure \ref{Fig:bindercrossBx1plane} shows an example of a Binder
crossing which can be used to locate the transition temperatures when
$J'/J$ is not too large.  For the $J'/J=0.5$ value shown, the crossing
is well-defined and occurs at $T_c/J \approx 1.12$.  For comparison,
when the frustrating interaction and transverse field vanish ($J'/J=0$
and $B_x=0$) we have $T_c/J=2.269$.

We can track $T_c$ from such Binder crossings only to $J'/J \approx
0.6$.  The reason is that, as with the $B_x=0$ case discussed by
Stephenson, a disorder line where the structure factor peak shifts from
${\bf q}=(\pi,\pi)$ to incommensurate values approaches the phase
transition line.  The evolution of this incommensuration shift is seen
in Fig.~{\ref{Fig:Sqshift}.  A peak in $S({\bf q})$ occurs at ${\bf
q}=(\pi,\pi)$ up to $J'/J \approx 0.7$.  At this point the peak moves
away from the $Neel$ value and evolves continuously towards the
triangular-lattice ${\bf q}=(2\pi/3,2\pi/3)$, as
shown in Fig.~\ref{Fig:Sqshift}.

Figure~\ref{Fig:conjecturedphasediag} shows the analog of Stephenson's
classical Ising model disorder line, which demarks the switch from
incommensurate peak in $S({\bf q})$ to a commensurate AF peak, in the
case of nonzero transverse field $B_x/J=1$.
It provides further insight into the failure of the Binder crossing
procedure, which worked at $J'/J=0.5$ as seen in 
Fig.~\ref{Fig:bindercrossBx1plane}.  The phase above $T_c$ for
$J'/J > 0.6$ no longer has a peak in $S({\bf q})$ at 
${\bf q}=(\pi,\pi)$, complicating the Binder scaling analysis.

\begin{figure}[t] 
\epsfig{figure=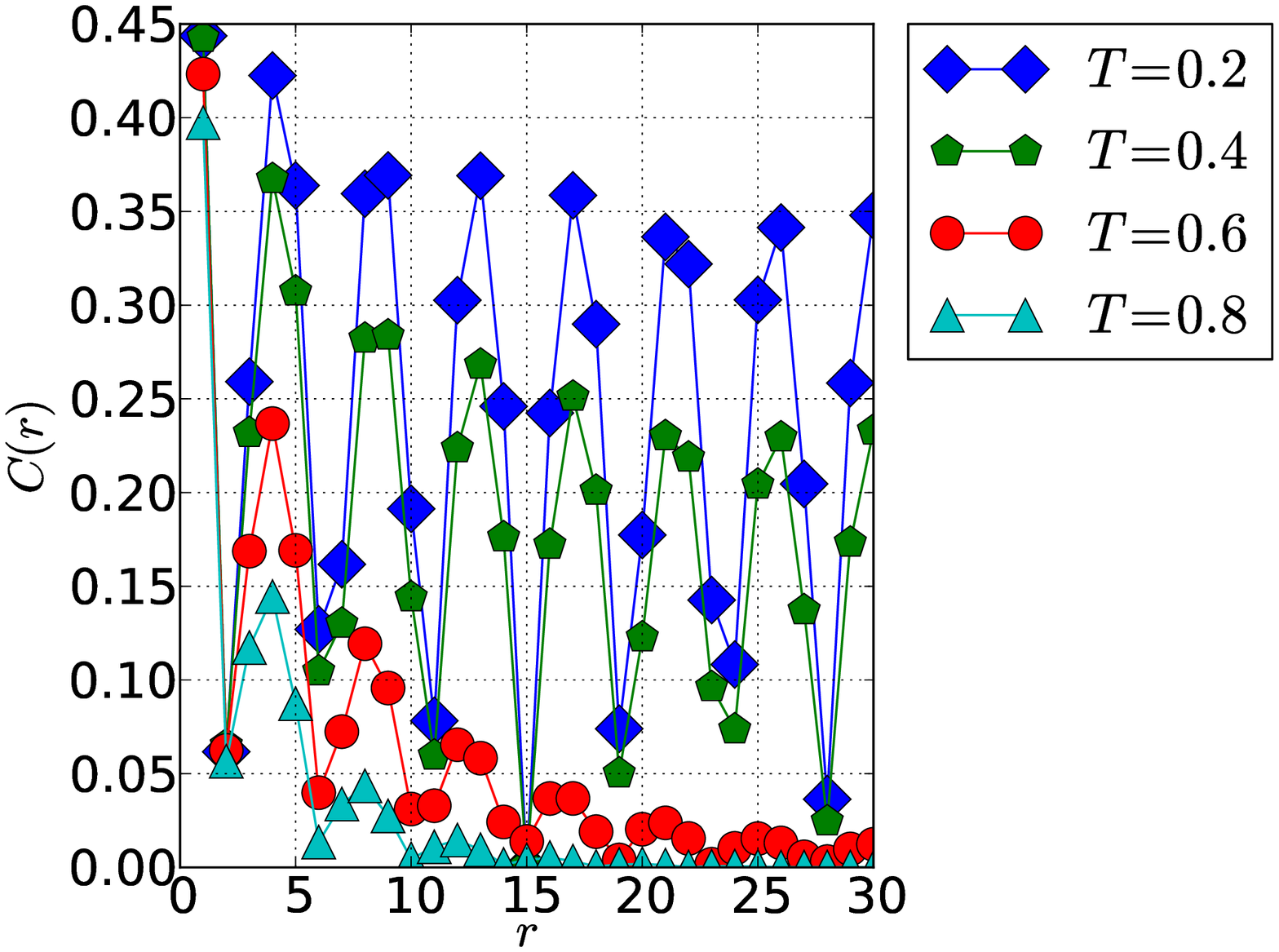,height=6cm,width=8.5cm,angle=0,clip}
\epsfig{figure=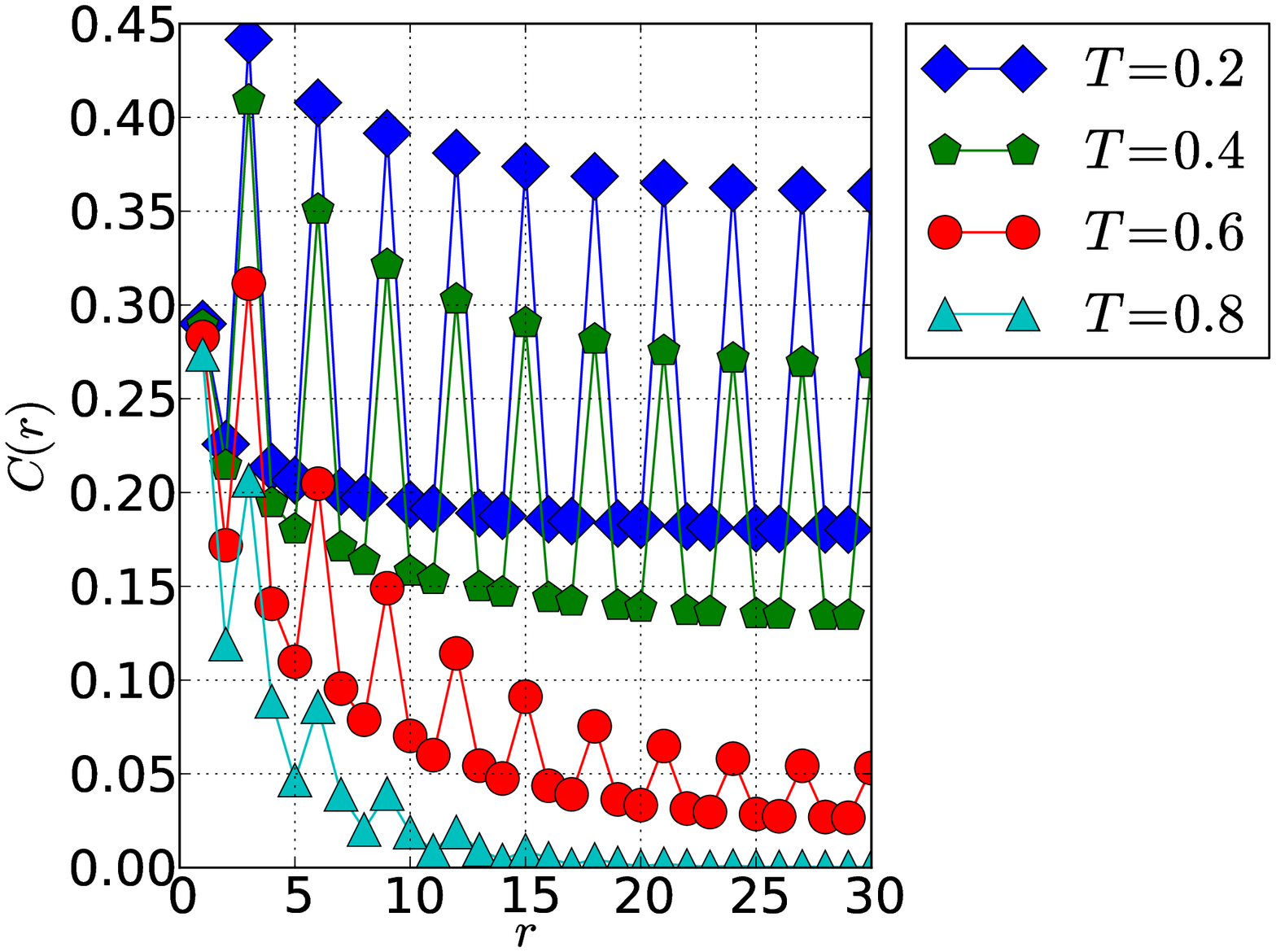,height=6cm,width=8.5cm,angle=0,clip}
\caption{(Color online)  
Correlation function $C(r)$ as a function of $r$ for several
temperatures and transverse field $B_x/J=1$.
\underbar{Upper Panel:  $J'/J=0.8$}:
The evolution of $C({\bf r})$ with $r$ exhibits a clear qualitative
difference for $T/J \leq 0.4$ and
$T/J \geq 0.6$, and is consistent with power law decay in the
lower temperature range, and an exponential decay at the
higher temperatures.  $C({\bf r}) \sim r^{-1/4}$ (with
$r=|{\bf r}|$) at $T/J \approx 0.4$
which is the expected universal power law value at the Kosterlitz-Thouless
critical temperature.
The oscillations in $C({\bf r})$ are associated with the fact the
structure factor peak is not at ${\bf q}=(\pi,\pi)$.
\underbar{Lower Panel:  $J'/J=1.0$}:  Same as upper panel
except in the triangular lattice limit.  Oscillations
are now at wavevector
${\bf q}=(2\pi/3,2\pi/3)$.
\label{Fig:CorLen_vs_r_B_1_Jp_0_8}
}
\end{figure}

\begin{figure}[t] 
\epsfig{figure=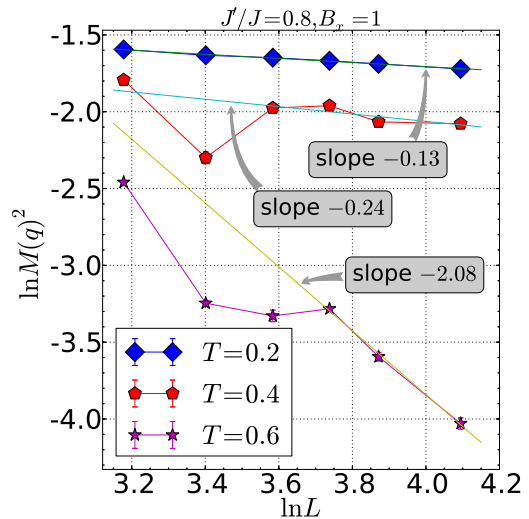,height=7.0cm,width=7.0cm,angle=0,clip}
\caption{(Color online)  
Below $T_{\rm KT}$ the structure factor decays as a power law 
with the lattice size, giving a straight line on a log-log plot, 
as shown occurs here for $T=0.2$ and $T=0.4$ at $J'/J=0.8$
and $B_x=1.0$. At higher $T = 0.6$ the slope simply reflects the lattice normalization in Eq.~\ref{Eq:Sq}.
\label{Fig:Mq_Jp_0_8_B_1_LnM2_vs_L}
}
\end{figure}

\begin{figure}[t] 
\epsfig{figure=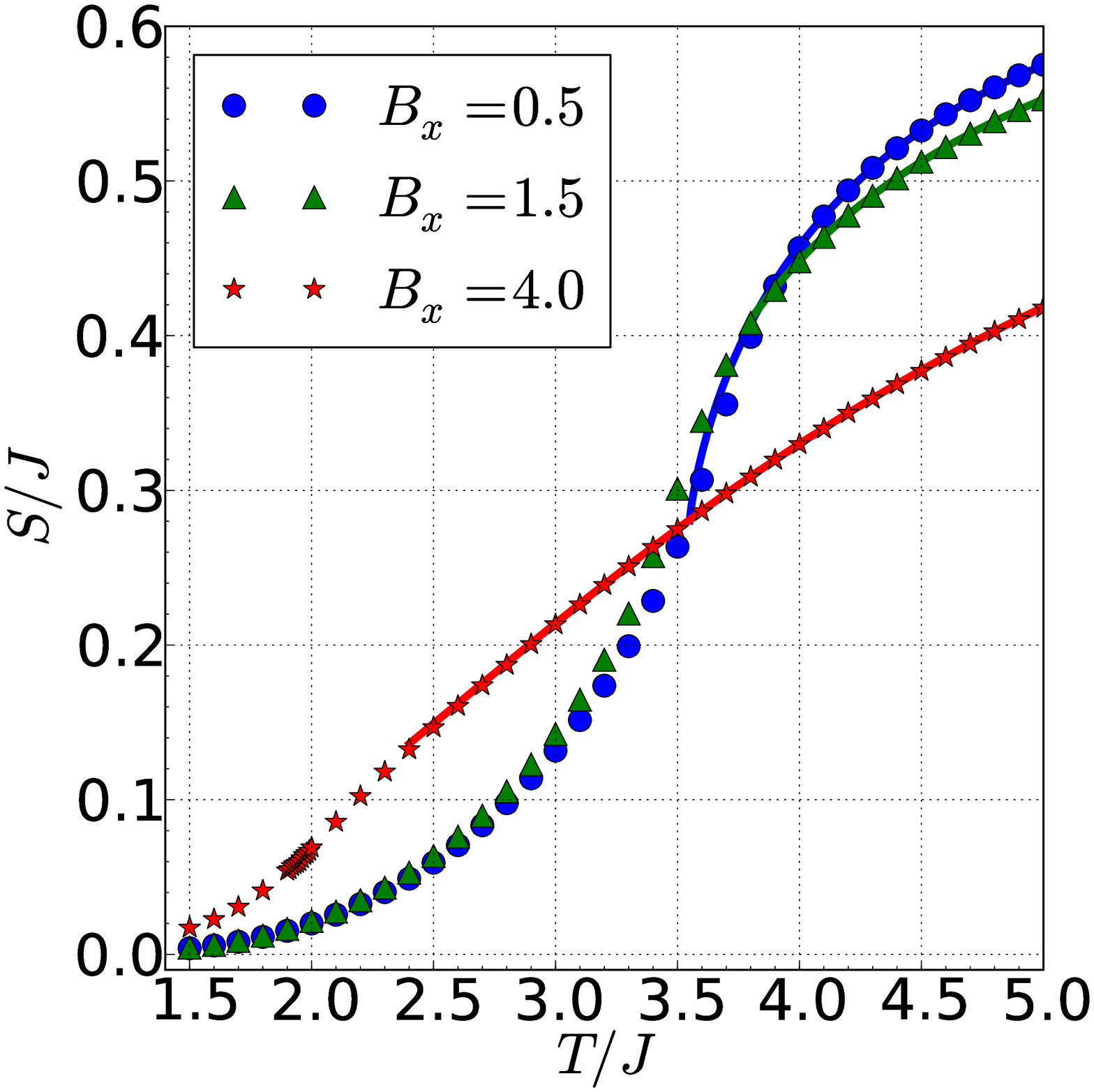,height=5.0cm,width=5.0cm,angle=0,clip}
\epsfig{figure=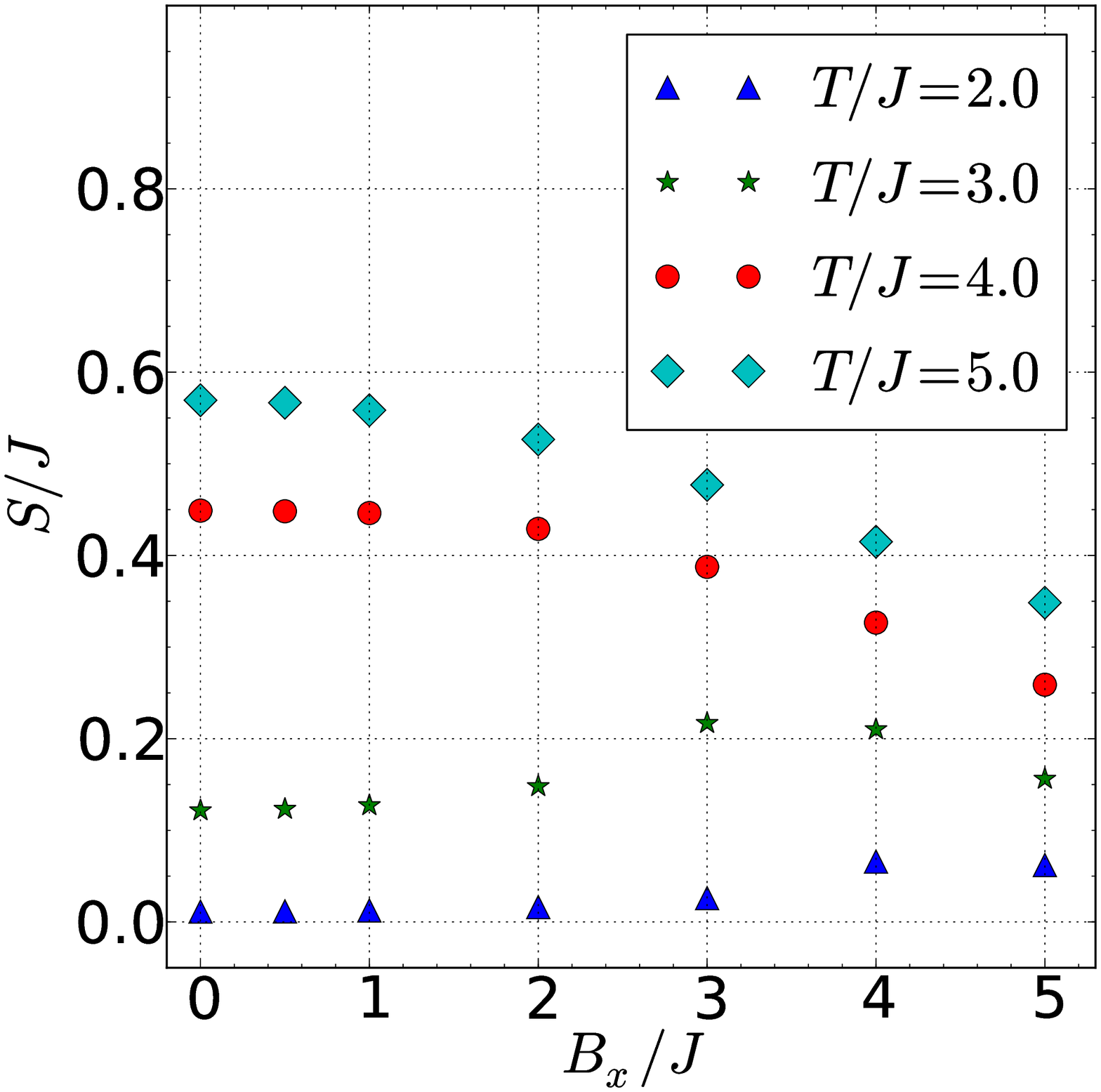,height=5.0cm,width=5.0cm,angle=0,clip}
\caption{(Color online)  
(a)  Entropy $s$ versus temperature $T$ for several fixed values of the
transverse field $B_x$.
(b)  Entropy $s$ versus transverse field $B_x$ for several fixed values of the
temperature $T$.
Squares (circles) are the results of CTQMC simulations on 27x27 lattice.
Solid curves are series expansions.
\label{Fig:SvsTandSvsBferro}
}
\end{figure}

The interplay of this commensurate-incommensurate transition with
Isakov-Moessner's observation of two phase transitions in the
triangular-lattice limit ($J'/J=1$) with the upper one being a
Kosterlitz-Thouless transition remains a tricky one.  Our suggested
phase diagram is given in Fig.~\ref{Fig:conjecturedphasediag},
where we show a dashed line connecting
a multicritical point to the upper phase transition found by Isakov and
Moessner. The lower transition may simply form a dome in the phase
diagram near the triangular-lattice limit. As the magnitude of $B_x$ is reduced, the multicritical point
will move closer to $J'/J=1$ eventually ending at the highly degenerate
point of the $T=0$ triangular-lattice Ising model.

Figures \ref{Fig:CorLen_vs_r_B_1_Jp_0_8}
and
~\ref{Fig:Mq_Jp_0_8_B_1_LnM2_vs_L}
provide numerical evidence for the Kosterlitz-Thouless region
in the interval $0.7 < J'/J < 1.0$ where the disorder line
crosses the phase transition line.
Figure \ref{Fig:CorLen_vs_r_B_1_Jp_0_8}
shows the spatial decay of the real space
correlation function $C({\bf r})$ for several temperatures
at $J'/J=0.8$.  It is clear that at $T/J=0.2$ and $0.4$ that 
$C({\bf r})$ is decaying much more slowly than at $T'/J= 0.6$ and higher.
While this change is
suggestive of a phase transition to a long-range ordered phase, Fig.~\ref{Fig:Mq_Jp_0_8_B_1_LnM2_vs_L}
puts things on a firmer footing.  Here a log-log plot of the
square of the structure factor versus linear lattice size 
gives the linear behavior which defines the 
Kosterlitz-Thouless phase
for $T=0.2$ and $T=0.4$.  The latter temperature has the
power law slope $-1/4$ expected at $T_{\rm KT}$.  At $T=0.6$ the largest
lattice sizes exhibit a decay with slope $-2$, a value 
generated by the normalization by $N=L^2$ in Eq.~\ref{Eq:Sq},
since the spatial sum gives a contribution which is a lattice size
independent constant at high temperatures owing to the exponential
fall-off of $C({\bf r})$.

%%%%%%%%%%%%%%%%%%%%%%%%%%%%%%%%%%%%%%%%%%%%%%%%%%%%%%%%%%%%%%%%%%
\section{Isentropes on the Nearest Neighbor Triangular Lattice}
%%%%%%%%%%%%%%%%%%%%%%%%%%%%%%%%%%%%%%%%%%%%%%%%%%%%%%%%%%%%%%%%%%

\subsection{Ferromagnetic Case}
We begin by showing in Fig.~\ref{Fig:SvsTandSvsBferro}(a) the entropy
function for the ferromagnetic nearest-neighbor Ising model as a
function of temperature and three values of transverse field. 
In this section, we will
let $J$ denote the magnitude of the nearest-neighbor interaction. 
Fig.~\ref{Fig:SvsTandSvsBferro}(b) shows the entropy function as a
function of $B_x/J$ at four fixed temperature values. In both cases the
symbols represent the results of the QMC simulations, while the solid
lines give the high temperature series expansion results. The heat
capacity as a function of temperature is shown in
Fig.~\ref{Fig:CvsTferro} for the same $B_x$ as the entropy plot. One
can see a dramatic supression in the specific heat peak as the quantum
critical point $B_{xc}=4/67$ is approached. 
In the thermodynamic limit, this must
imply a sharp reduction in the amplitude for the specific heat
divergence caused by the substantial loss of entropy before the
transition to long-range order occurs.\cite{heistrisq,hubtrisq} 

In Fig.~\ref{Fig:PDandisentropesferro}
we show the phase diagram
for the model together with the isentropic contours. While there is a
very healthy amount of entropy in the system at the transition in small
transverse fields (it is approximately 46 percent of the total entropy),
when $B_x/J$ exceeds $4$ the entropy at the transition is very small-
less than 10 percent of the total entropy. The quantum critical point is
known from previous studies to be at $B_x/J=4.67$.\cite{oitmaa,reiger}
We found that, as the quantum critical point is approached and the
transition temperature $T_c$ goes to zero, the entropy along the
transition vanishes as $T_c^2$. This is consistent with general scaling
arguments.\cite{qimiao}
The appearance of minima in the isentropes near the phase transition
line has been seen in a number of other classical and quantum models
\cite{cone11}.

\begin{figure}[t] 
\epsfig{figure=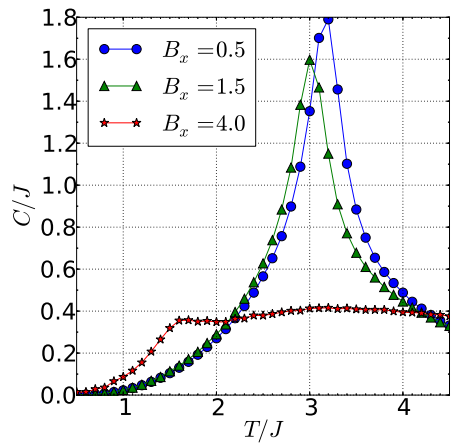,height=6.0cm,width=6.0cm,angle=0,clip}
\caption{(Color online)  
Specific heat $C(T)$ for the ferromagnetic nearest-neighbor model.
The quantum fluctuations introduced by the transverse field $B_x$ 
reduce $T_c$ from the classical triangular lattice $T_c=3.64$ until,
for sufficiently large $B_x$, order no longer occurs at any finite $T$.
\label{Fig:CvsTferro}
}
\end{figure}

\begin{figure}[t] 
\epsfig{figure=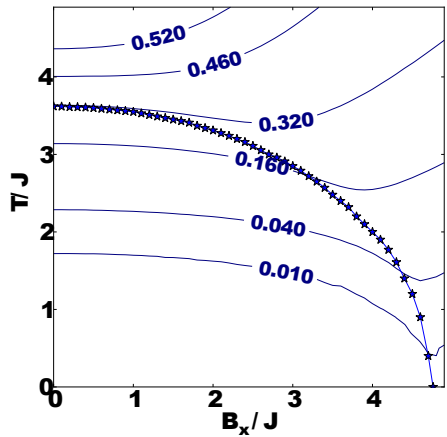,height=6.0cm,width=6.0cm,angle=0,clip}
\caption{(Color online)  
Phase diagrams and isentropes of the ferromagnetic triangular lattice
model in the temperature ($T/J$) vs megnetic field ($B_x/J$) plane.
Isentropes obtained from CTQMC on a 27x27 lattice and high temperature
expansions. Phase boundary is obtained from the Binder cumulant. 
\label{Fig:PDandisentropesferro}
}
\end{figure}

\vfill\eject

\subsection{Antiferromagnetic nearest-neighbor triangular lattice model}

We now turn to the antiferromagnetic nearest-neighbor Ising model on the
triangular lattice in a transverse field. The entropy as a function of
temperature at fixed $B_x$ and as a function of $B_x$ at fixed
temperatures are shown in 
Fig.~\ref{Fig:SvsTandSvsBantiferro}.
The heat capacity for a few selected
$B_x$ values as a function of temperatures is shown in 
Fig.~\ref{Fig:CvsTantiferro}.
Note that, in this case, the peak in the heat capacity is
associated with short-range order. Any long range order occurs only at
much lower temperatures. 

Fig.~\ref{Fig:PDandisentropesantiferro}
shows the phase diagram and isentropes
of the antiferromagnetic model. Here, the phase boundaries showing two
transitions as a function of temperature are taken from the work of
Isakov and Moessner. Note that the triangular lattice Ising model has a
substantial ground state entropy, but that must be removed at $T=0$ with
quantum fluctuations. This means that all contours of constant entropy
between the values of $0.320$ and zero must originate from the $T=0$,
$B_x=0$ point. That singular limit is difficult to approach numerically.

We see that a magnetic field of $B_x/J$ slightly less than unity leads
to the largest transition temperature. But, at the transition the
entropy is rather small- only about one tenth of the total entropy in the
system. This shows that these phase transitions involve a very small
fraction of the states and may not be easy to get to.

\begin{figure}[t] 
\epsfig{figure=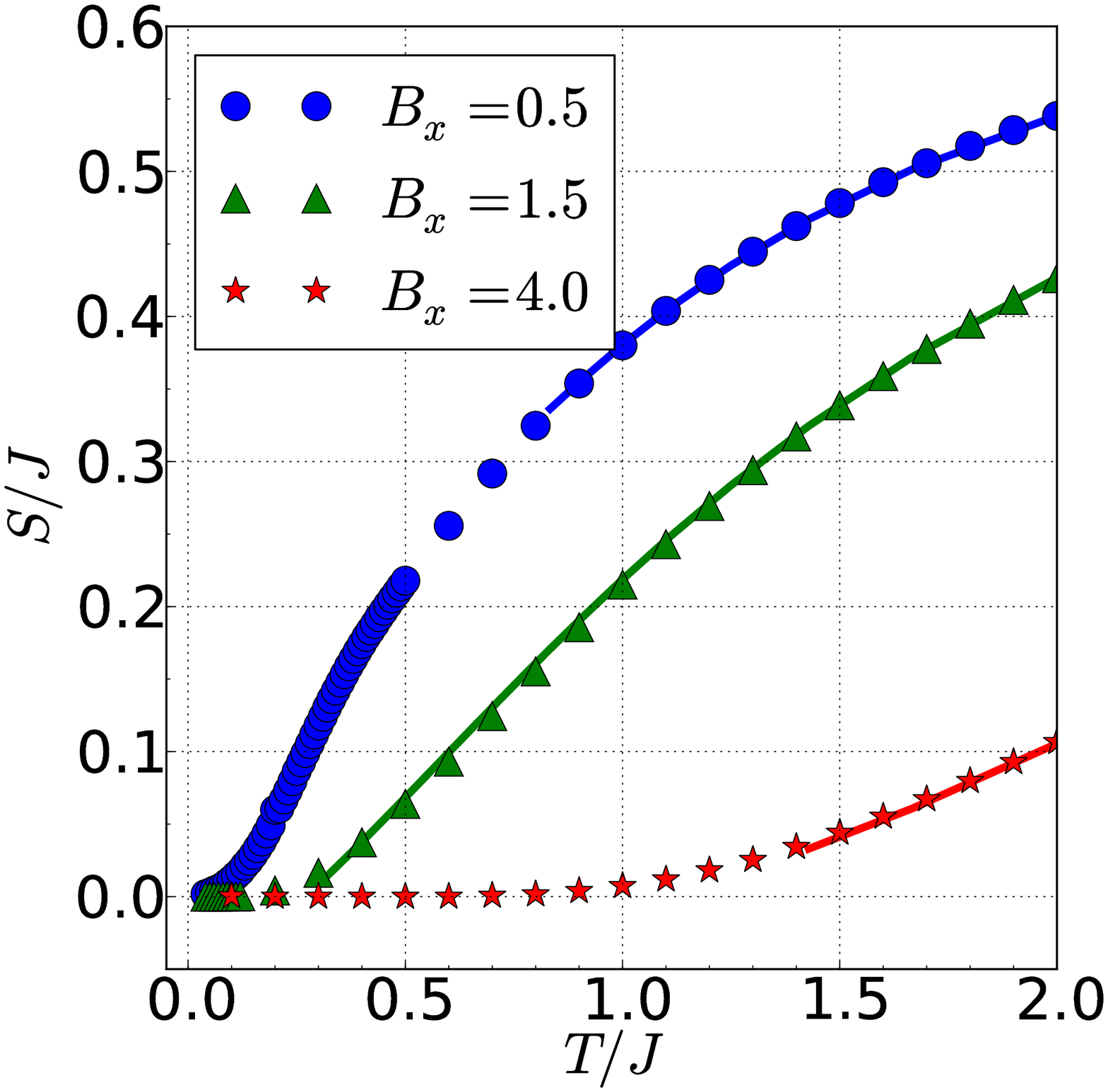,height=6.0cm,width=6.0cm,angle=0,clip}
\epsfig{figure=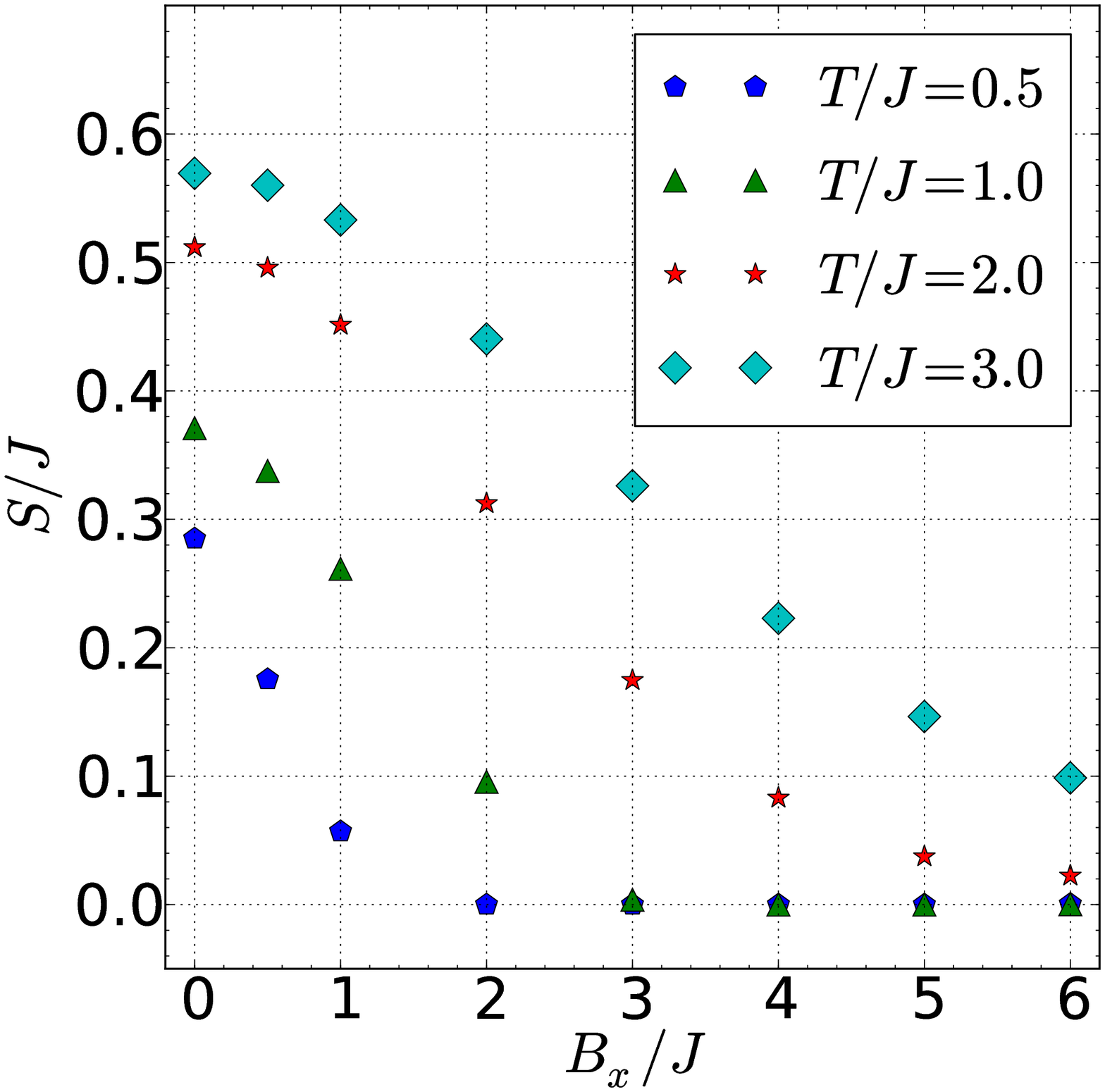,height=6.0cm,width=6.0cm,angle=0,clip}
\caption{(Color online)  
(a)  Entropy $s$ versus temperature $T$ for several fixed values of the
transverse field $B_x$.
(b)  Entropy $s$ versus transverse field $B_x$ for several fixed values of the
temperature $T$.
Squares (circles) are the results of CTQMC simulations on a 30x30 lattice.
Solid curves are series expansions.
\label{Fig:SvsTandSvsBantiferro}
}
\end{figure}

\begin{figure}[t] 
\epsfig{figure=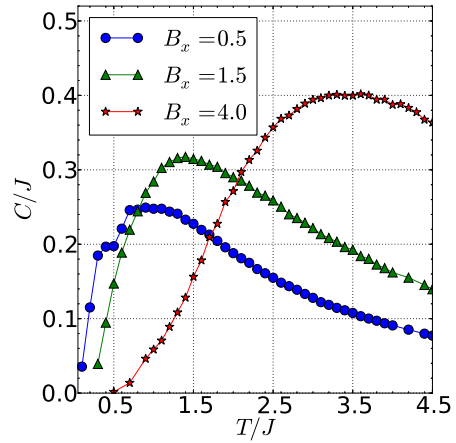,height=6.0cm,width=6.0cm,angle=0,clip}
\caption{(Color online)  
Specific heat $C(T)$ for the antiferromagnetic case.  The peak in 
$C(T)$ is much broader than when $J<0$, but shows a similar suppression 
towards $T=0$ as $B_x$ grows.
\label{Fig:CvsTantiferro}
}
\end{figure}

\begin{figure}[t] 
\epsfig{figure=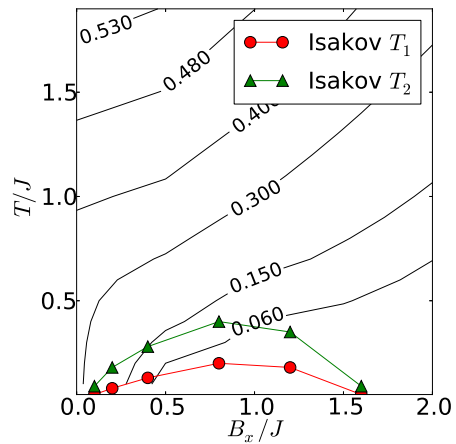,height=6.0cm,width=6.0cm,angle=0,clip}
\caption{(Color online)  
Phase diagram and isentropes of the near neighbor antiferromagnetic case obtained from CTQMC on the 30x30 lattice in the $T/J$ vs $B_x / J$ plane and high temperature expansions. Phase diagram was obtained by Isakov and Moessner.
}
\label{Fig:PDandisentropesantiferro}
\end{figure}

%%%%%%%%%%%%%%%%%%%%%%%%%%%%%%%%%%%%%%%%%%%%%%%%%%%%%%%%%%%%%%%%%%
\subsection{Effect of Longer Range Interactions}
%%%%%%%%%%%%%%%%%%%%%%%%%%%%%%%%%%%%%%%%%%%%%%%%%%%%%%%%%%%%%%%%%%

In the NIST experiments\cite{britton12}, spin-1/2 Be ions in a Penning
trap interact with long range interactions which fall off with a
tunable power law $J_{ij} \propto 1/r_{ij}^a$ with $0 < a < 3$.  This
range of functional forms allows quantum simulations going from infinite
range ($a=0$) to Coulomb ($a=1$), to dipole ($a=3$).  Because the
interactions are antiferromagnetic as well as long-range, the
experiments correspond to `large-scale' frustration, and, ultimately, it
is hoped they will allow the realization of associated novel (eg.~spin
liquid) phases.  In this section we will extend our numerical work on
the thermodynamics of the nearest-neighbor antiferromagnetic transverse
field Ising model on a triangular lattice to include several
realizations of longer ranged interactions.  While we will not study
systems with $J_{ij}$ non-zero between all pairs, these simulations give
first indications of the evolutions of the entropy in situations with
large-scale frustration.

The isentropes with a non-zero next-nearest neighbors interaction $J'$
are shown in Fig.~\ref{Fig:PDandisentropesantiferrolongrange}.  
At low
transverse fields this changes the isentropes as the tremendous ground
state degeneracy of the nearest-neighbor Ising model is now lifted.
However, at and above a transverse field of $B_x/J\approx 1$, the
isentropes are very similar to the nearest-neighbor case and 
phase transitions would only occur at very low values of the entropy.

\begin{figure}[t] 
\epsfig{figure=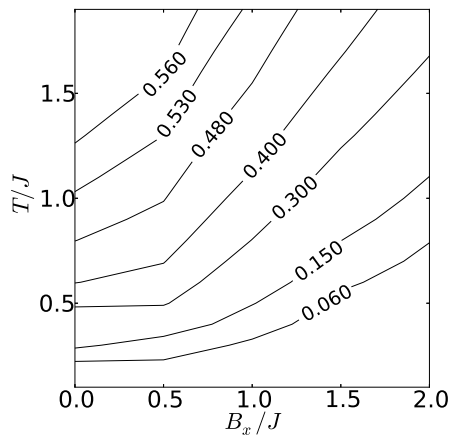,height=6.0cm,width=6.0cm,angle=0,clip}
\caption{
Isentropes of the antiferromagnetic case with non-zero next near
neighbor interaction $J'=J / \sqrt{3}^{(1.4)}$.  The change from 
Fig.~\ref{Fig:PDandisentropesantiferro} with $J'=0$ is minimal.
In particular, the key message is that the entropy values remain low 
in the region where phase transitions might occur.
\label{Fig:PDandisentropesantiferrolongrange}
}
\end{figure}

%%%%%%%%%%%%%%%%%%%%%%%%%%%%%%%%%%%%%%%%%%%%%%%%%%%%%%%%%%%%%%%%%%
\section{Conclusions}
%%%%%%%%%%%%%%%%%%%%%%%%%%%%%%%%%%%%%%%%%%%%%%%%%%%%%%%%%%%%%%%%%%

In this paper we have provided quantitative results for the
thermodynamics and phase transitions of the Ising model in a transverse
field, with an emphasis on frustration and the effect of longer range
interactions. For the isotropic triangular-lattice model, we have studied the entropy
function for both ferromagnetic and antiferromagnetic exchange interactions. Using Quantum
Monte Carlo simulations together with high temperature expansions and high
field expansions, we have obtained the contours of constant entropy in the
temperature, transverse-field plane. Our main conclusion is that phase
transitions in the frustrated antiferromagnetic 
model occur at very low entropy compared to the unfrustrated case and these
results are not substantially altered by adding weaker further neighbor exchange
interactions.

We have presented a study of the quantum Ising model on the anisotropic
triangular lattice that interpolates between the square and triangular
lattice limits that yields new results for the phase diagram and
thermodynamic properties. This model, with frustrating antiferromagnetic
interactions, raises many subtle issues and presents a real challenge to
any computational study.  We first obtained the disorder line where the
short-range order moves away from ($\pi,\pi$). Unlike the $B_x=0$ case
studied by Stephenson, in presence of a transverse-field, the ground
state order does not stay at ($\pi,\pi$) all the way from the
square-lattice ($J'=0$) to the triangular-lattice ($J'=J$) limit.  The
disorder line crosses the phase transition line near $J'/J=0.7$ and
heads sharply towards $T=0$. While it represents a crossover from
short-range commensurate to short-range incommensurate order above this
crossing, it presumably becomes a phase transition line separating long
range commensurate order from power-law incommensurate order. At $T=0$,
it should turn into a Lifshitz point marking the onset of incommensurate
long-range order. 

Once the system has incommensurate order, the Binder ratios no longer
show size independent crossings and it is difficult to determine the
transition temperature for the expected Kosterlitz-Thouless phase.
Rather than study many $J'/J$ values, we focused on $J'/J=0.8$. We found
that as the temperature was reduced there was a sharp change in the
nature of spin-spin correlation functions between $T=0.6 J$ and $T=0.2
J$. The correlations are very small at large distances and consistent
with exponential decay at $T=0.6 J$, whereas at $T=0.2 J$ they remain
substantial at the largest distances accessible to our simulations and
only a close inspection shows a power-law decay.  A peak in the structure
factor also moves with the size of the system making it difficult to
precisely locate the transition temperature. At $T=0.4 J$, the
correlation functions are consistent with a $r^{-1/4}$ decay, and hence
we take this to be the transition temperature. At very low temperatures,
the system may lock into other commensurate phases, as it does for the
triangular-lattice case through higher order anisotropies, but those are
beyond our simulation capabilities.

Cold atom experiments on the triangular lattice, transverse field Ising model
are ongoing\cite{britton12}.  Our results provide quantitative data
on the entropies required to reach possible phase transitions for
different values of $B_x$.
In addition to the connections of the present work to 
these experiments, and to triangular-square
lattice interpolations of the Heisenberg model\cite{zheng99}, similar
experimental
studies are now being undertaken on itinerant electron magnetism as
realized in Hubbard models \cite{tarruel12}, as well as
companion theoretical treatments\cite{bluemer13}. The focus thus far has been
on Dirac points on honeycomb lattices and topological features in the
band structure.  However, the behavior  of spin correlations on such
tunable lattices is one of the key goals of the next generation of
experiments.

%\section*{ACKNOWLEDGMENTS}

This work was supported by the University of California UCLAB program
and by the NNSA under DE-NA0001842-0 and by the 
National Science Foundation grant number DMR-1004231.

%%%%%%%%%%%%%%%%%%%%%%%%%%%%%%%%%%%%%%%%%%%%%%%%%%%%%%%%%%%%%%%%%%%%%%%%
%%%
%%%%%     BIBLIOGRAPHY
%%%%%%%%%%%%%%%%%%%%%%%%%%%%%%%%%%%%%%%%%%%%%%%%%%%%%%%%%%%%%%%%%%%%%%%%%%%


\begin{thebibliography}{100}

\bibitem{schultz64}
T.D. Schultz and D.C. Mattis, 
Rev. Mod. Phys. {\bf 36}, 856 (1964).

\bibitem{pfeuty70}
P. Pfeuty,
Annals of Physics {\bf 57}, 79 (1970).

\bibitem{suzuki71}
M. Suzuki, Prog. Theor. Phys. {\bf 46}, 1337 (1971); and
Prog. Theor. Phys. {\bf 56}, 2454 (1976).

\bibitem{stinchcombe73}
R.B. Stinchcombe,
J. Phys. {\bf C6}, 2459 (1973).

\bibitem{blote02}
H.W.J. Bl\"ote and Youjin Deng,
Phys. Rev. {\bf E66}, 066110 (2002).

\bibitem{sachdev-book}
S. Sachdev, {\it Quantum Phase Transitions},
Cambridge University Press, Cambridge, United Kingdom (1999).

\bibitem{chandra-sondhi-moessner}
R. Moessner, S. L. Sondhi, and P. Chandra, Phys. Rev. Lett. 84, 4457 (2000).

\bibitem{wannier50}
G.H. Wannier, Phys. Rev. {\bf 79}, 357 (1950).

\bibitem{isakov03}
S.V. Isakov and R. Moessner,
Phys. Rev. {\bf B68}, 104409 (2003).

\bibitem{degennes63}
P.G. de Gennes, Solid St. Comm. {\bf 1}, 132 (1963).

\bibitem{stinchcombenote}
Reference \onlinecite{stinchcombe73} contains an early list of 
systems related to the Ising model in a transverse field.

\bibitem{monroe10}
%% ``Quantum Simulation and Phase Diagram of the Transverse Field Ising Model
%%  with Three Atomic Spins," 
E.E. Edwards, S. Korenblit, K. Kim, R. Islam,
M.-S. Chang, J.K. Freericks, G.-D. Lin, L.-M. Duan, and C. Monroe,
Phys. Rev. {\bf B82}, 060412 (2010);
%% ``Quantum Simulation of the Transverse Ising Model with Trapped Ions," 
K. Kim, S. Korenblit, R. Islam, E. E. Edwards, M.-S. Chang, C. Noh, H.
Carmichael, G.-D.Lin, L.-M. Duan, C.-C. Joseph Wang, J.K. Freericks,
and C. Monroe, New J. Physics {\bf 13}, 105003 (2011); and
%% ``Onset of a Quantum Phase Transition with a Trapped Ion Quantum
%% Simulator," 
R. Islam, E. E. Edwards, K. Kim, S. Korenblit, C. Noh, H.
Carmichael, G.-D.Lin, L.-M. Duan, C.-C. Joseph Wang, J. K. Freericks,
and C. Monroe, Nature Communications {\bf 2}, 377 (2011).

\bibitem{britton12}
%% ``Engineered two-dimensional Ising interactions in a trapped-ion quantum
%% simulator with hundreds of spins,"
J.W. Britton, B.C. Sawyer, A.C. Keith, C.-C. Joseph Wang,
J.K. Freericks, H. Uys, M.J. Biercuk and J.J. Bollinger
Nature {\bf 484}, 489 (2012).

\bibitem{tabei08}
S.M.A. Tabei, F. Vernay, and M.J.P. Gingras,
Phys. Rev. {\bf B77}, 014432 (2008);
S.M.A. Tabei, M.J.P. Gingras, Y.-J. Kao, and T. Yavors'kii,
Phys. Rev. {\bf B78}, 184408 (2008).

\bibitem{LiHoYF}
For a recent comprehensive review see A. Dutta, U. Divakaran, D. Sen, B. K. Chakrabarti, T. F. Rosenbaum, G. Aeppli, cond-mat:arXiv:1012.0653.

\bibitem{book}
 J. Oitmaa, C. Hamer and W-H. Zheng, {\it Series Expansion
Methods for strongly interacting lattice models}, (Cambridge
University Press 2006).

\bibitem{series-review}
M.P. Gelfand, and R.R.P. Singh,
{\it High-order convergent expansions for quantum many particle systems},
Advances in Physics, {\bf 49} N1:93-140 (2000).

\bibitem{sandvik} A. W. Sandvik, Phys. Rev. {\bf E68}, 056701 (2003).

\bibitem{zheng99}
See for example, Zheng Weihong, Ross H. McKenzie, and Rajiv R. P. Singh,
Phys. Rev. {\bf B59}, 14367 (1999).

\bibitem{selke}
W. Selke and L. N. Shchur, Phys. Rev. {\bf E80}, 042104 (2009).

\bibitem{stephenson70}
J. Stephenson,
J. Math. Phys. {\bf 5}, 1009 (1964); {\it ibid.} {\bf 11}, 420 (1970);
{\it ibid.} {\bf 11}, 413 (1970).

\bibitem{eggarter75} 
T.P. Eggarter, 
Phys. Rev. {\bf B12}, 1933 (1975).

\bibitem{bak} 
P. Bak, Rep. Prog. Phys. {\bf 45}, 587 (1982).

\bibitem{selke88} W. Selke, Physics Reports {\bf 170}, 213 (1988).

\bibitem{oitmaa}
J. Oitmaa, C. J. Hamer and Z. Weihong, J. Phys. A: Math. Gen. {\bf 24}, 
2863 (1991);
H-X He, C. J. Hamer and J. Oitmaa, J. Phys. A: Math. Gen. {\bf 23}, 1775 (1990).

\bibitem{reiger}
H. Reiger and N. Kawashima, Eur. Phys. Jour. {\bf B9}, 233 (1999).


\bibitem{qimiao}
L. Zhu, M. Garst, A. Rosch and Q. Si, Phys. Rev. Lett.
{\bf 91}, 066404 (2003).

\bibitem{heistrisq}
P. Sengupta, A. W. Sandvik and R. R. P. Singh,
Phys. Rev. {\bf B68}, 094423 (2003).

\bibitem{hubtrisq}
J. Merino and R. H. Mckenzie, Phys. Rev. Lett. {\bf 87}, 237002 (2001).

\bibitem{cone11}
J.D. Cone, A. Zujev and R.T. Scalettar,
Phys. Rev. {\bf B83}, 045108 (2011).

\bibitem{tarruel12}
L. Tarruell, D. Greif, T. Uehlinger, G. Jotzu, and T. Esslinger,
Nature {\bf 483}, 10871 (2012).

\bibitem{bluemer13}
N. Bluemer, C-C. Chang, and R.T. Scalettar,
unpublished.

\end{thebibliography}
\end{document}